\documentstyle[preprint,eqsecnum,aps]{revtex}

\begin{document}
\draft
\title{Two interacting particles in an effective 2-3--d
random potential }
\author{Fausto Borgonovi\\}
\address{
Dipartimento di Matematica, Universit\`a Cattolica,
via Trieste 17, 25121 Brescia, Italy  \\
Istituto Nazionale di Fisica Nucleare, Sezione di Pavia,
via Bassi 6, 27100 Pavia, Italy.
}
\author {Dima L. Shepelyansky$^{*}$}
\address {
Laboratoire de Physique Quantique, Universit\'{e} Paul Sabatier,\\
118, route de Narbonne, 31062 Toulouse, France
}

\date{\today}
\maketitle
\begin{abstract}
We study the effect of coherent propagation of two interacting particles
in an effective 2-3-d  disordered potential.
Our numerical data demonstrate that in dimension
$d > 2$,  interaction can lead to
two--particles
delocalization  below one--particle delocalization border.
We also find that the distance between the two
delocalized particles (pair size)
grows logarithmically with time.
As a result pair propagation is subdiffusive.
\end{abstract}

\pacs{
\hspace{2.9cm}
}


\newpage
\section{Introduction}
\label{sec:level1}

The question of interacting particles in a random potential has
recently got a great deal of attention (see for example \cite{Kirk}).
Indeed  this problem is important for the understanding of
conduction of  electrons in metals and disordered systems.
It is also very interesting from the theoretical view point
since it allows  to understand the effects of interaction on
Anderson localization.
It is a common belief that in one-dimensional (1d) systems
near the ground state,
a repulsive  interaction between particles leads to a  stronger localization
if compared with  non interacting case \cite{Tieri}.
Even if more complicate, the 2--dimensional  (2d) problem
is assumed to be localized, while in
the  3--dimensional (3d) case
 delocalization can take place in the presence
 of interaction \cite{Kirk}. However this  problem is rather
difficult  for analytical, experimental and numerical investigations
and therefore it is quite far from its final resolution.

The complicated nature of the above problem can be illustrated
by the example of only two interacting particles (TIP) in  a
random potential. Indeed in this case, contrary to the common
lore, even  repulsive particles can create an effective
pair which is able to propagate on a distance $l_c$
much larger than its own size of the order of
one particle localization length $l_1$ \cite{TIP}.
{}From one side  interference effects for non--interacting
particles force the particles to stay together
at a distance $\sim l_1$ even
in the repulsive case. But the relative motion between
the two particles leads to the destruction of such interference
and allows their  coherent propagation on a distance $l_c \gg l_1$.
More explicitly, according to \cite{TIP},
in the quasi 1d case with $M$ transverse
channels one has
\begin{equation}
{l_{c} \over {l_1}} \sim {l_1 M} { U^2\over {32 V^2}}
\label{est}
\end{equation}
where $U$ is the strength of on site interaction and $V$ is the
one particle hopping matrix element. Here the  inter-site distance
is $a=1$ and the wave vector $k_F \sim 1$.
Since $\l_1 \propto M$, then $l_c \propto M^3$.
In 2d
the effective number of channels due to
2d localization  is $M \sim l_1$
so that $l_c \sim l_1^3$ and localization is preserved
(here and everywhere $l_1$ represents the one--particle
localization length in any dimension).
The sharp increase of $l_c$ with the number of transverse channels $M$ leads
to a straightforward possibility
of delocalization for a pair of particles in 3d
while one particle
remains localized \cite{Dreply,Imry,nonli}.
In this sense the 3d case
is much more interesting
due to the possibility of delocalization
and we qualitatively discuss it below.

One of the interesting features of pair delocalization in 3d is
that it is not  due to a simple shift of the mobility edge produced
by  the interaction (such possibility is indeed
not so interesting). In fact, it is possible to consider a system in which
{\it all} one-particle eigenstates are localized for {\it all}
energies. This can be
for example the 3d Lloyd model with diagonal disorder
$E_{n_1,n_2,n_3} = \tan \phi_{n_1,n_2,n_3}$ and hopping $V$ on a cubic lattice,
where $\phi_{n_1,n_2,n_3}$ are random phases homogeneously distributed in the
interval $[0,\pi]$. In this model {\it all} one-particle eigenstates are
localized for $V < V_{c} \sim 0.2$. However, two repulsive
particles with on-site
interaction $U$ can create a coupled state which is delocalized
and   propagates through the lattice.
Indeed, a pair "feels" only smoothed potential \cite{TIP}
that corresponds to an effective renormalization of
the  hopping matrix element
$V_{eff}$ which is strongly enhanced due to interaction and becomes
larger than $V_c$. Of course, the enhancement takes place only
for sufficiently large one-particle localization length $l_1 \gg 1$.
Therefore, the hopping $V<V_c$ should be not far from $V_c$
although there is, a priori, no requirement  for $V$ to be very close
(parametrically) to $V_c$.

The two particles delocalization due to interaction
takes place only for states in which particles are on a distance $R<l_1$
from each other while for $R \gg l_1$ eigenstates are localized.
Such  kind of situation is quite unusual since it means that the
absolutely continuous spectrum of the Schr\"{o}dinger operator,
corresponding to the delocalized pair, is {\it embedded} into
the pure point
spectrum of localized almost noninteracting particles states.
For $R \gg l_1$ the interaction between the two particles
is exponentially small and this implies a very
small coupling between the
states corresponding to these two kinds of spectra.
However, due to the quasi degeneracy of levels, even
a small coupling can lead to important modifications
of the above picture as it was discussed in \cite{nonli}.
Therefore, a direct numerical investigation of
interaction-assisted delocalization in 3d is highly desirable.
While the recent theoretical arguments and
numerical simulations in quasi--1d case
\cite{TIP} - \cite{oppen}
definitly demostrate the existence of enhancement
for $l_c$ no numerical simulations have been done
in 3d case. Indeed, in 3d  basis grows as $N^6$,
where $N$ is the number of 1d unperturbed one-particle states,
and that leads to heavy numerical problems.

A similar type of interaction-assisted
delocalization can
be also realized  in the kicked rotator model (KRM) \cite{KRM}
in 3d. In this case the unitary evolution
operator takes the place of Schr\"{o}dinger operator
and eigenenergies are replaced by quasi-energies.
The advantage of such models
is due to the independence of localization length on quasi-energy so
that all one-particle states in 3d are localized for
$V<V_c$ and delocalized for $V>V_c$.
Even if very efficient, numerical simulations for
KRM in 3d become very difficult;
for two particles  situation becomes even worse
due to $N^6$ basis growth.

One of the ways to overcome
these numerical difficulties is the following.
For 1d KRM the number of dimensions can be effectively
modelled by introducing  a frequency modulation of the perturbation
parameter \cite{Dima,1and}.
The case with  $\nu$  incommensurate frequencies in the kick modulation
corresponds  to an effective solid state model with
dimension $d=\nu+1$.  For $\nu=2$, the effective dimension
$d=3$ and Anderson transition can be efficiently investigated
\cite{1and}.
A similar approach can be done for two interacting particles and
it allows to gain a factor $N^4$ in numerical simulations.

In this paper we investigate the model of two interacting kicked rotators
(KR) studied in \cite{TIP}, \cite{nonli}
with frequency modulation and $\nu=2,3$.
The quantum dynamics is described by the  evolution operator~:
\begin{equation}
\begin{array}{c}
{\hat S_2} = \exp \{ -i [
 H_{0}({\hat n})+H_{0}({\hat n'})+U\delta_{n,n'}] \} \\
\times \exp \{-i [ V(\theta,t) + V(\theta',t) ]\}
\end{array}
\label{qmap}
\end{equation}
with ${\hat n}^{(')}=-i {\partial}/{\partial {\theta^{(')}}}$.
Here $H_{0}({ n})$ is a random function of $n$ in the interval
$[0,2\pi]$ and it describes the unperturbed spectrum of rotational phases.
The perturbation $V$ gives the coupling between the unperturbed levels
and has the form
$V(\theta,t)= k(1+\epsilon \cos\theta_1 \cos\theta_2 \cos\theta_3) \cos\theta$
with $\theta_{1,2,3}=\omega_{1,2,3} \;t$.
In the case of two modulational frequencies
($\nu=2, \omega_3=0$), as in \cite{1and},
we choose   frequencies $\omega_{1,2}$  to be
incommensurate with each other and with the frequency $2\pi$
of the kicks. Following \cite{1and} we take
$\omega_1=2\pi\lambda^{-1}$, $\omega_2=2\pi\lambda^{-2}$ with
$\lambda=1.3247...$ the real root of the cubic equation
$x^3-x-1=0$. For $\nu=3$ we used the same $\omega_{1,2}$ and
$\omega_{3} = 2\pi/{\sqrt{2}}$. We also studied another
case of functional dependence of $V(\theta)$
analogous to \cite{1and} and corresponding to the Lloyd model.
All computations have been done for symmetric configurations.
According to the theoretical arguments \cite{TIP} and numerical
simulations \cite{oppen} the antisymmetric configurations corresponding to
fermions with nearby site interaction should show a similar type of behaviour.

The paper is constructed as follows. In  section II we
discuss the model and present the main results for $\nu=2$.
The case of $\nu=3$ is discussed in  section III.
The kicked rotator model corresponding to the 3d
Lloyd model is studied in section IV.
Conclusions and discussions of results are presented in section V.

\section{The KRM model with two frequencies}
\label{sec:level2}

Before to discuss the effects of interaction let us first discuss the
noninteracting case $U=0$. Here the evolution operator can be
presented as a product of two operators $S_1$ describing the
independent propagation of each particle:
\begin{equation}
\begin{array}{c}
{\hat S_1} = \exp ( -i  H_{0}({\hat n})) \exp (-i V(\theta,t))
\end{array}
\label{qmap1}
\end{equation}
Since $V$ depends on time in a quasiperiodic way with
$V(\theta,t)=V(\theta, \theta_1, \theta_2)$ and
$\theta_{1,2} = \omega_{1,2} t$, one can go to
the extended phase space \cite{Dima}, \cite{1and} with effective
dimension $d=3$. In this space the operator is independent on time and
has the form
\begin{equation}
\begin{array}{c}
{\hat {S_1}^{~}} = \exp ( -i  H_{1}({\hat n, \hat n_1, \hat n_2}) )
\exp (-i V(\theta,\theta_1,\theta_2))
\end{array}
\label{exsp1}
\end{equation}
with $H_{1}(n, n_1, n_2) = H_0 (n)+\omega_1 n_1+\omega_2 n_2$.
Due to linearity in $n_{1,2}$ the transformation from (\ref{qmap1}) to
(\ref{exsp1}) is exact. However, the numerical simulations of
(\ref{qmap1}) are $N^2$ times more effective than for (\ref{exsp1}).

The system (\ref{exsp1}) corresponds to
an effective 3d model. Numerical simulations in \cite{1and}
showed that the variation of coupling amplitude $V$ gives
the transition from localized to diffusive regime as in usual
Anderson transition in $3d$. In \cite{1and} the form
of the kick $V$ had been chosen as
\begin{equation}
\begin{array}{c}
V(\theta,\theta_1,\theta_2)=
-2\tan^{-1}[2k(\cos\theta+\cos\theta_1+\cos\theta_2)-E]
\end{array}
\label{LL}
\end{equation}
In this case after a mapping
similar to the one used in \cite{FISH} the equation for
eigenfunction with a quasi-energy $\mu$ can be presented in a usual
solid-state form:
\begin{equation}
\begin{array}{c}
T_{\bf n} u_{\bf n} +k
{\sum_{\bf r}} u_{\bf n - r} =
E u_{\bf n}
\end{array}
\label{TT}
\end{equation}
where the sum is taken only over nearby sites and
$T_{\bf n}=\tan((H_1(n,n_1,n_2)-\mu)/2)$,
${\bf n} = (n, n_1, n_2) $. For random phases under tangent
 the diagonal disorder is distributed in Lorentzian way
and the model becomes equivalent to the 3d Lloyd model.
While we also investigated the kick form (\ref{LL})
(see section IV) our main results have been obtained for
\begin{equation}
\begin{array}{c}
V(\theta,\theta_1,\theta_2)=k\cos \theta (1+\epsilon \cos\theta_1 \cos\theta_2)
\end{array}
\label{VKR}
\end{equation}
in the case of two frequencies $\nu=2$. According to \cite{D87}
in this case the equation for eigenfunctions can be also reduced
to an effective solid-state Hamiltonian which, however, has a bit more
complicated form than (\ref{TT}). We choose (\ref{VKR})
since it was numerically more efficient than (\ref{LL}).
To decrease the number of parameters we always kept $\epsilon=0.75$.

The one-particle transition as a function of coupling (hopping)
parameter $k$ in (\ref{VKR}) is presented in Fig.1.
Similar to \cite{1and} the localization length $l_1$ is
determined from the stationary probability distribution over
unperturbed levels ${\vert \psi_n \vert}^2 \sim
\exp(-2 \vert n \vert/l_1)$ while the diffusion rate is extracted
from the gaussian form  of the probability distribution
$\ln W_n \sim -n^2/2Dt$ with
$D=<n^2>/t$. According to Fig.1 the transition takes place
at the critical hopping value $k_{cr} \approx 1.8$.
Below $k_{cr}$ all quasi-energy states are localized.
The independence of transition point from quasi-energy
is one of useful properties of KR models.

Our main aim was the investigation of TIP effects well below the
transition point $k_{cr}$. As in \cite{nonli}
we characterized the dymamics (\ref{qmap}) by the second moments along
the diagonal line $n=n'$~:
$
\sigma_{+} (t) = \langle ( \vert n\vert +
\vert n' \vert )^2 \rangle_t  /4
$
and across it
$
\sigma_{-} (t) = \langle ( \vert n \vert - \vert n' \vert )^2 \rangle_t
$.
We also computed the total probability distribution along and
across this diagonal \cite{nonli}: $P_{\pm}(n_{\pm})$
with $n_{\pm} = \vert n \pm n' \vert /2^{1/2}$. The typical case
is presented in Figs. 2,3.  These pictures definitely show the
appearence of pair propagation even if the interaction is neither
attractive nor repulsive. Indeed, the pair size
is much less than the distance on which two particles are
propagating together. If we fit the probability distibution in Fig.3
as $P_{\pm} \sim \exp(-2n_{\pm}/l^{\pm})$ then we can see
that the ratio $l^+/l^- \sim 25$ ($l_c \approx l^+ \approx 95$) is quite large.
It is interesting to note that $P_+(n_+)$ at different moments of time
is closer to an exponential ($\ln P_+ \sim n_+$) than to a gaussian
($\ln P_+ \sim {n_+}^2$). The spreading along the lattice
leads only to growth of $l^+$ with time but the shape of distribution does
not corresponds to a diffusive process.

Another interesting feature of Fig.2 is the slow decrease of the rate
of $\sigma_+$ growth  and the slow growth of $\sigma_-$. To check if
the growth of $\sigma_+$ is completely suppressed with time
we analysed its dependence on $t$ for different values of interaction $U$
(Fig.4, probability distribution is shown
in Fig.5). For $U \leq 0.5$ the growth of $\sigma_+$ is
completely suppressed while for $U=1$ the complete suppression
is a bit less evident. To understand in a better way the case
$U=1$ we can look on the dependence of the number of effectively
excited levels $\Delta N$ on time. To estimate $\Delta N$ we should
rewrite (\ref{qmap}) in the extended basis where the evolution operator
has the form:
\begin{equation}
\begin{array}{c}
{\hat {S_2}^{~}} = \exp ( -i [ H_{0}({\hat n})+H_{0}({\hat n'})
+\omega_1 {\hat n_1} +\omega_2 {\hat n_2}+U\delta_{n,n'}] )
\exp (-i [V(\theta,\theta_1,\theta_2)+V(\theta',\theta_1,\theta_2)])
\end{array}
\label{exsp2}
\end{equation}
Since $\Delta n_{1,2} \approx \Delta n^{(')} \approx \Delta n_+$
the number of excited levels
can be estimated as $\Delta N \approx
\Delta n_+ \Delta n_- \Delta n_1 \Delta n_2
\approx {\sigma_+}^{3/2}  {\sigma_-}^{1/2}$.
Following the standard estimate  based on the
uncertainty relation  \cite{D87}
a delocalization can take place only if $\Delta N$ grows
faster than the first power of $t$. As our numerical data show
( see  Fig. 6)  the ratio $W=\Delta N /t$ remains approximately constant
or is even  slightly decreasing in time. This indicates that our
case is similar to localization in $2d$ where this ratio also
remains constant for very long time.
The reason why below $k_{cr}$ the situation is similar to
$2d$ can be understand in the following way.
According to (\ref{exsp2}) the total dimension
is 4 and we have there 2 particles. Therefore, we can argue
that the dimension per particle is 2 and that  below
the 3d delocalization border $k_{cr}$ our system effectively
represents two particles in an effective
dimension $d_{eff}=2$. However, two particles in 2d
are always localized but the localization length can be
exponentially large. The dependence of $\sigma_+$ at fixed moment of
time on $k$ is presented in Fig.7 and indeed, it demonstrates
a sharp increase of $\sigma_+$ with $l_1$ approaching
$k_{cr}$. Therefore, we conclude that for $k<k_{cr}$ our model
effectively represents TIP in 2d. Below $k_{cr}$ the pair created
by interaction remains localized but the localization length $l_c$
grows exponentially with $l_1$. To see effects of interaction
in $d_{eff} > 2$ we should study the system with three modulational
frequencies $\nu=3$. But before to analyse the case
$\nu=3$ we would like to discuss the behaviour of $\sigma_-$.

Indeed, Figs. 2,4  clearly demonstrate a slow growth of $\sigma_-$
with time which means the increase of the size of the pair $\kappa$.
The results presented in Fig.8 show that $\kappa \approx {\sigma_-}^{1/2}$
grows logarithmically with time as $\kappa \approx C_L \ln t$ where
$C_L$ is some time independent factor being $C_L \approx 0.8 (U=2)$
and $C_L \approx 0.6 (U=1)$. Of course, this logarithmic growth should
terminate after the complete localization in $\sigma_+$
but this time scale $t_c$ is very large and for $t < t_c$
we have clear logarithmic growth of $\kappa$. As discussed in \cite{nonli}
we attribute this growth to the fact that propagating in a random potential
the pair is affected by some effective noise which leads to a slow separation
of two particles. Indeed, the matrix elements of interaction $U_-$
decay exponentially fast with the growth of the pair size $n_- = \kappa$
according to a rough estimate
$U_- \sim U_s\exp(- \vert n_- \vert /l_1)$ with $U_s \sim U/{l_1}^{3/2}$.
These small but finite matrix elements lead to the growth of the
pair size $\kappa$ with slow diffusion rate $D_- \propto {U_s}^2
\exp(- 2\vert n_- \vert /l_1)$. According to the relation
$\kappa^2/t \approx D_-$ the pair size grows as $\kappa \sim l_1\ln t/2 $
\cite{nonli} which  is in agreement with data of Fig.7.
More detailed numerical
simulations are required to verify the dependence $C_L \sim l_1$.

\section{The KRM model with three frequencies}
\label{sec:level3}

According to the above discussion the suppression of diffusive
growth of $\sigma_+$ can be explained by two factors. The first one
is that the effective dimension is $d_{eff}=2$ and localization always
takes place in 2d. Another reason is the slow logarithmic growth of
pair size.
To separate these two effects we investigated the dynamics of
TIP in the KRM with three modulational frequencies $\nu=3$. In the extended
phase space the evolution operator has the form
\begin{equation}
\begin{array}{c}
{\hat {S_2}^{~}} = \exp ( -i [ H_{0}({\hat n})+H_{0}({\hat n'})
+\omega_1 {\hat n_1} +\omega_2 {\hat n_2}+\omega_3 {\hat n_3}
+ U\delta_{n,n'}] ) \\
\exp (-i [V(\theta,\theta_1,\theta_2,\theta_3)+
V(\theta',\theta_1,\theta_2,\theta_3)])
\end{array}
\label{exsp3}
\end{equation}
with
\begin{equation}
\begin{array}{c}
V(\theta,\theta_1,\theta_2)=k\cos \theta
(1+\epsilon \cos\theta_1 \cos\theta_2 \cos\theta_3)
\end{array}
\label{VKR3}
\end{equation}
For $U=0$ we have one particle in 4d and transition to delocalization
takes place above a critical value of perturbation parameter $k_{cr}$.
According to our numerical data $k_{cr} \approx 1.15$
for $\epsilon = 0.9$ (Fig.9). Below $k_{cr}$ all eigenstates are exponentially
localized. For $U \neq 0$ the total dimension in (\ref{exsp3}) is 5 and
since we have 2 particles the effective dimension per particle
is $d_{eff} =5/2$. Since $d_{eff} > 2$ the first argument given above
becomes not relevant and we expect TIP delocalization below $k_{cr}$.
Let us to note that above $k_{cr}$ the above TIP problem becomes
not interesting since even without interaction the particles
spread along the lattice and the interaction between them does not
affect significantly their dynamics.

The numerical simulations of TIP for (\ref{exsp3}) - (\ref{VKR3})
in one-particle localized phase $k < k_{cr}$ demonstrate strong enhancement
of two particles propagation. A typical case is presented in Figs. 10,11.
According to these data the growth of $\sigma_+$ is unlimited
and TIP delocalization takes place below $k_{cr}$. The analysis
of $\sigma_-(t)$ shows that pair size
$\kappa \approx {\sigma_-}^{1/2}$  grows logarithmically with
time similar to the case with $\nu=2$. We think that this slow growth
of $\kappa$ is responsable for a slow decrease of the pair diffusion rate
$D_+=\sigma_+/t$ with time. Due to the increase of $\kappa$
the probability to have a distance between particles
of the order of $l_1$ decreases as
$1/\kappa(t) \sim 2/(l_1\ln t)$ and therefore, we expect that the diffusion
rate
of the pair will decrease with time as $D_+ \sim D_{ef}/{\ln}^\mu t$.
Here $\mu=1$ and $D_{ef}$ is some effective "subdiffusion" rate.
While the above probability argument gives $\mu=1$ it is quite possible
that sticking in the region with $\kappa \gg l_1$ will give a
faster decrease of $D_+$ with a higher value of $\mu$.
As it was discussed in \cite{nonli}  the growth of
pair size should also give logarithmic corrections to the
coherent localization length in the quasi-1d case
(\ref{est}) ($l_c \sim {l_1}^2/\ln^{\mu} l_1$).

Another confirmation for the delocalization transition below
one--particle threshold is given by the analysis of the
number of effectively excited states. Indeed for $\nu=3$
one has $\Delta N \approx \sigma_+^2 \sigma_-^{1/2}$.
According to our data, for sufficiently strong interaction $U$
the quantity $W=\Delta N/t$ grows approximately linearly with time
(see Fig. 12) while for small $U$ values $W$ decreases with time.
Contrary to the case  $\nu=2$ this indicates that TIP delocalization
takes place  for $U$ values  bigger than a critical
$U_{cr} \approx 0.7$. Above this critical value the number of
excited states at a given time grows when $k$ approaches
to one--particle delocalization border $k_{cr}$ (Fig. 13).

\section{The effective Lloyd model}
\label{sec:level4}
We also studied the model  (\ref{qmap}) with the kick perturbation
given by (\ref{LL}). In this case, the non interacting problem
can be reduced to the Lloyd model with pseudo--random
sites energies \cite{1and}. For $\nu=2$ and $E=0$
the one particle delocalization
border is $k_{cr} \approx 0.46$ \cite{1and}. For TIP problem
the behaviour of this model is similar to that of section II.
The strong enhancement of propagation is demonstrated in Fig. 14.
Even if the investigation of this model is more difficult for
the numerical simulations, our data indicate, as it was in section II,
that for $\nu=2$ the suppression of $\sigma_+$ is always present.
In the same way  we attribute this behaviour
 to the effective two dimensionality of the model $d_{eff} =2$.

We also analyzed the tangent model (\ref{qmap}), (\ref{LL})
for the case of three frequencies $\nu=3$
($ V(\theta,\theta_1,\theta_2,\theta_3) =
-2 \tan^-1 (2 k (\cos\theta +\cos\theta_1 +\cos\theta_2
+\cos\theta_3) ) $).
This case is similar to that discussed in section III.
The behaviour of $\sigma_{\pm}$ is presented in Fig. 15
and indicates the existence of delocalization transition
for TIP below one particle delocalization border with
$k_{cr} \approx 0.22$.

\section{Conclusions and discussions}
\label{sec:level5}
Our numerical investigations definitely demonstrate the effect
of enhancement of the localization length for TIP in a random
potential.
These results were obtained for kicked rotators models with
frequency modulation. Such approach allows to model efficiently
TIP problem in an effective dimension $d_{eff} \geq 2$.
Numerical data for these models confirm the theoretical
expectations \cite{Dreply,Imry,nonli}
that TIP delocalization in $d > 2$ is possible below
one--particle delocalization border.
In agreement with \cite{nonli} we found TIP pair delocalization
and, at the same time, a logarithmic growth of the pair size.
We attribute this growth to the noise produced by the random
potential. Indeed a pair propagating in a random potential
sees different realizations of disorder which act like
some effective noise. Such noise originates transitions which
increase the distance between the two particles.
Even if the amplitude of these transitions is  exponentially
decreasing  with the two--particle distance, it gives rise
to logarithmic growth of pair size with time.
This in turn produces a  subdiffusive pair propagation
$( \Delta n_+ )^2  \sim D_{ef} t/\ln^\mu t$.
We give arguments for $\mu=1$, but it is possible that due to
sticking in the region with large distance between particles
one can have $\mu > 1$.
Further investigations should be done in order to determine
the exact value of $\mu$.
Another qualitative argument for the subdiffusive propagation
is the tunneling between states in which the two particles are far
from each other (at a distance     $R \gg l_1$) and states
in which particles stay within $l_1$  ($R \leq l_1$).
These states are quasi-degenerate since the spectrum
of delocalized states is embedded in the spectrum of localized ones.
In this situation even an exponentially small overlapping between
these two kinds of states becomes important and it can lead to
a subdiffusive pair propagation.
Further work  should be done for a better understanding of the
final spectrum structure and the eigenfunctions properties for TIP
in 3d.

\section{Acknowledgments}
\label{sec:level6}
One of us (D.L.S) would like to thank the University of Como
for hospitality during the final stage of this work.

\begin{figure}
\caption{
One-particle Anderson transition in the model
(\protect\ref{qmap1})
with $V$ from
(\protect\ref{VKR})
$\nu=2$, $\epsilon = 0.75$ as a function of hopping $k$.
Critical point $k_{cr} \approx 1.8$. One particle
localization lengths $(l_1)$ and diffusion
coefficients ($D$) are evaluated from the fitting of  probability
distribution. Error bars indicate the standard deviation obtained from
an ensemble of $100$ (localized) and $10$ (diffusive) different
random realizations. Lines are drawn to fit an eye.
}
\end{figure}

\begin{figure}
\caption{
Dependence of second moments on time
in 2-particles model (\protect\ref{qmap}) with $V$ from
(\protect\ref{VKR}) and
$\nu=2$, $k=0.9$, $\epsilon=0.75$,;
upper curve is $\sigma_+$($ U=2$), middle is $\sigma_-$ ($ U=2$),
lower is $\sigma_+$  ($ U=0$). At $t=0$
both particles are at $n=n'=0$, basis is
$-250 \leq n,n' \leq 250$. Insert shows in larger scale the two lower curves
in the interval $8 \times 10^5 < t < 10^6$.
}
\end{figure}

\begin{figure}
\caption{
Probability distribution at $t=10^6$ as a function
of $n_\pm = 2^{-1/2} (n \pm n') $
for the case of Fig.2 with $U=2$:
$P_+ (n_+ )$ (full line); $P_- (n_- )$ (dashed); dotted line is the
distribution $P_+ (n_+ )$ for $U=0$.
}
\end{figure}

\begin{figure}
\caption{
Same as Fig.2 but for $U=1$.
}
\end{figure}

\begin{figure}
\caption{
Same as Fig.3 but for $U=1$.
}
\end{figure}

\begin{figure}
\caption{
 Dependence of $W=\Delta N/t ={\sigma_+}^{3/2} {\sigma_-}^{1/2}/t$ on time
for $\nu=2$, $k=0.9$, $\epsilon=0.75$
 and $U=2$ (full line), $U=1$ (dashed line),
$U=0.5$ (dotted line).
}
\end{figure}

\begin{figure}
\caption{
 Dependence of $\sigma_+$ on  $k$ for $\nu=2$ at $t=2 \times 10^5$,
 $ U=2 $ (full circles), $ U=0 $ (open circles).
}
\end{figure}

\begin{figure}
\caption{
Dependence of ${\sigma_{-}}^{1/2}$ on time $\ln t$
for cases of Fig.2 $(U=2, k=0.9)$ (full upper line) and
Fig.4 $(U=1, k=0.9)$ (dashed lower line).
}
\end{figure}

\begin{figure}
\caption{
One-particle Anderson transition in the model
(\protect\ref{qmap1}) with
$V$ from (\protect\ref{VKR3})
$\nu=3$, $\epsilon = 0.9$ as a function of hopping $k$.
Critical point $k_{cr} \approx 1.15$.
One particle localization lengths ($l_1$) and diffusion
coefficient ($D$) are evaluated from the fitting of  probability
distribution. Error bars indicate the standard deviation for
ensemble of $100$ (localized) and $10$ (diffusive) different
random realizations.
Lines are drawn to fit an eye.
}
\end{figure}
\begin{figure}
\caption{
Dependence of second moments on time
in model (\protect\ref{qmap}) with $V$ from (\protect\ref{VKR3}) and
$k=0.7$, $\epsilon=0.9$;
upper curve is $\sigma_+ $ ( $ U=2$ ), middle is $\sigma_-$ ($ U=2 $),
lower is $\sigma_+$ ( $ U=0 $). At $t=0$
both particles are at $n=n'=0$, basis is
$-250 \leq n,n' \leq 250$, $\nu=3$.  Inset shows the dependence
of ${\sigma_-}^{1/2}$ on $\ln(t)$.
}
\end{figure}
\begin{figure}
\caption{
Probability distribution at $t=1.6 \times 10^6$ as a function
of  $n_\pm = 2^{-1/2} (n \pm n') $
for the case of Fig. 10:
$P_+ (n_+ )$ (full line); $P_- (n_- )$ (dashed); dotted line is the
distribution $P_+ (n_+ )$ for $U=0$.
}
\end{figure}

\begin{figure}
\caption{
Dependence of $W=\Delta N/t = \sigma_+^2 \sigma_-^{1/2}/t$ on time
for $\nu =3$, $k=0.7$, $\epsilon=0.9$ and $U=2$ (upper curve),
$U=1$ (middle curve), $U=0.5$ (lower curve).
}
\end{figure}

\begin{figure}
\caption{
Dependence of $\sigma_+$ on  $k$ for $\nu=3$ at $t=2 \times 10^5$,
$U=2$ (full circles), $U=0$ (open circles).
}
\end{figure}

\begin{figure}
\caption{
Dependence of second moments on time
for the Lloyd model with $\nu=2$,
$k=0.35$, $U=2$;
upper curve is $\sigma_+$, lower is $\sigma_-$.
At $t=0$
both particles are at $n=n'=0$.  Basis is
$-256 < n,n' < 256 $. For $U=0$ one has
$\sigma_+ \approx 16$.
}
\end{figure}

\begin{figure}
\caption{
Dependence of second moments on time
for the Lloyd model with $\nu=3$,
$k=0.2$, $U=2$;
upper curve is $\sigma_+$, lower is $\sigma_-$.
At $t=0$
both particles are at $n=n'=0$. Basis is
$128 < n,n' < 128 $. For $U=0$ one has
$\sigma_+ \approx 2$.
}
\end{figure}

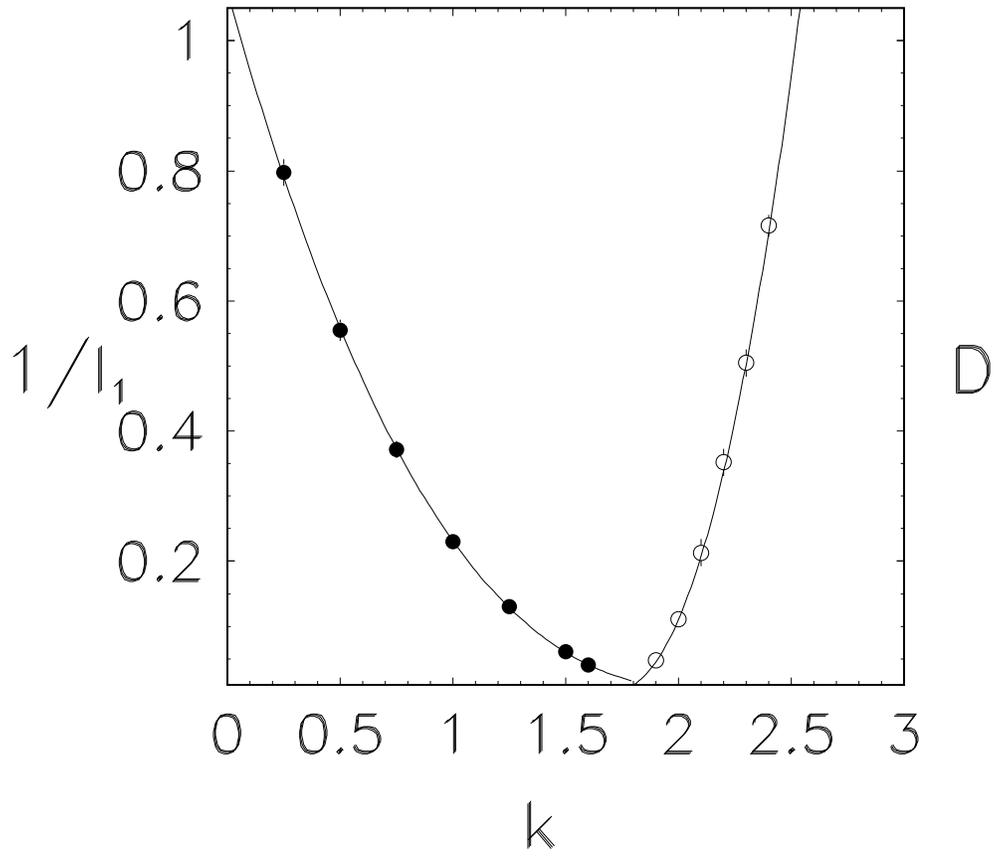
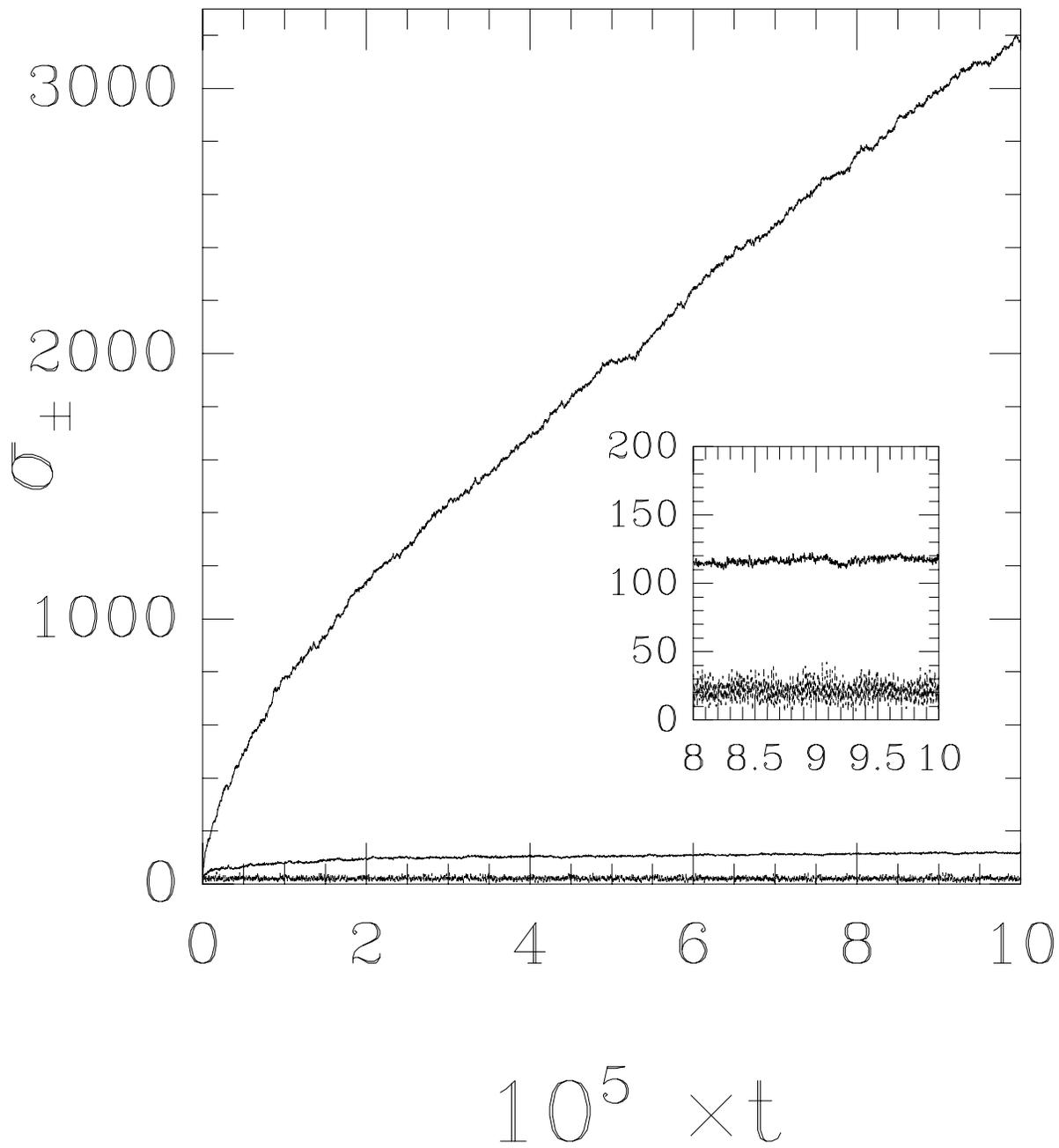
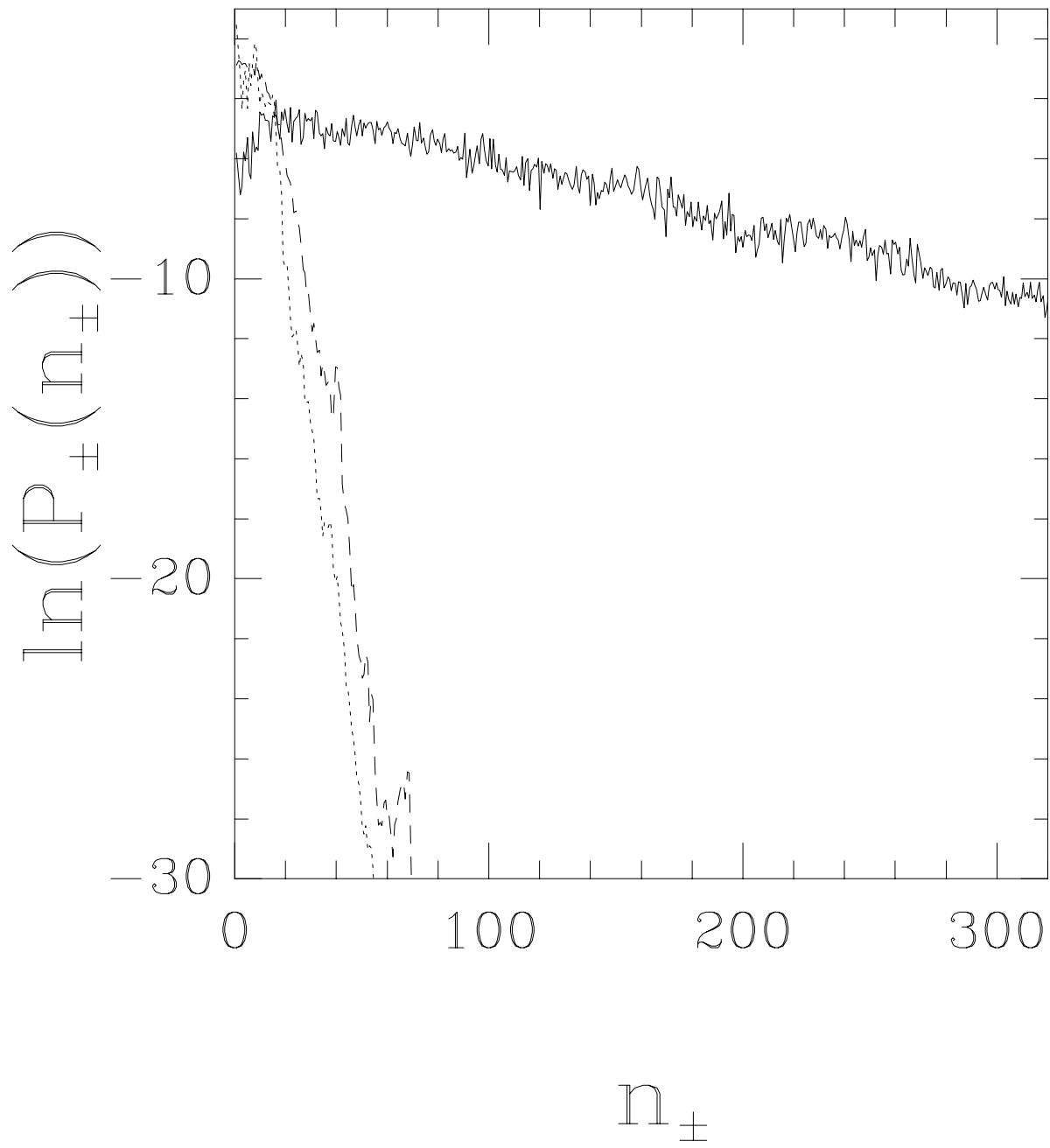
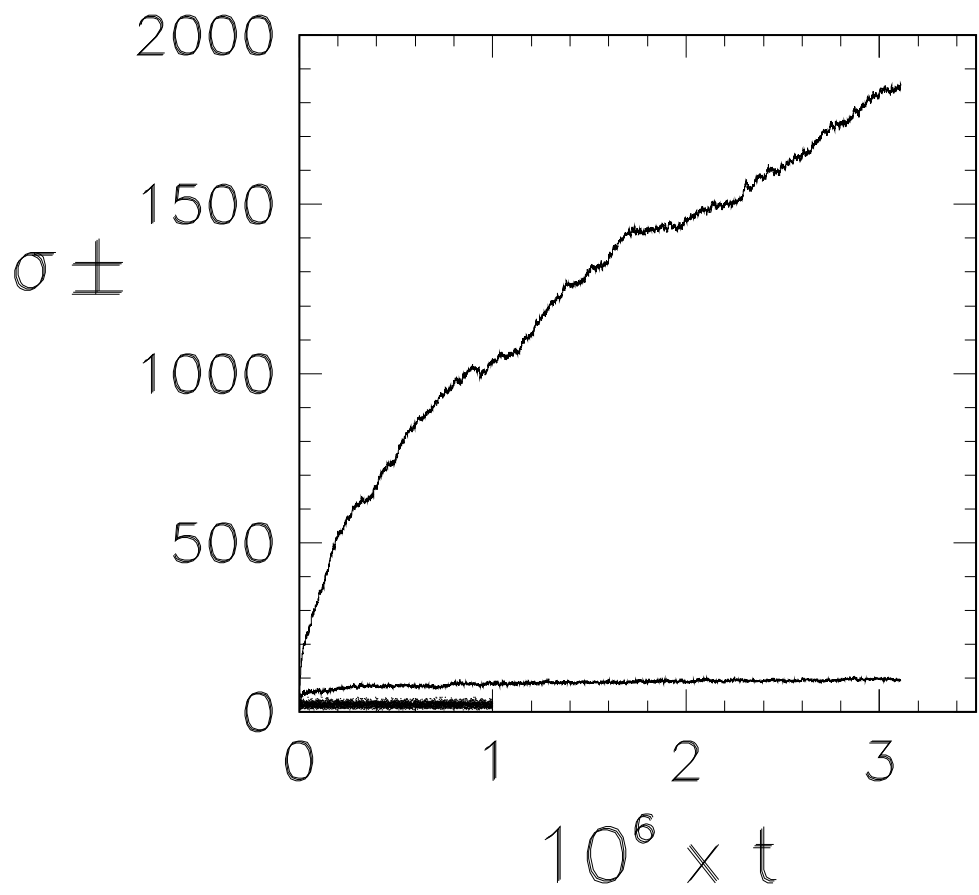
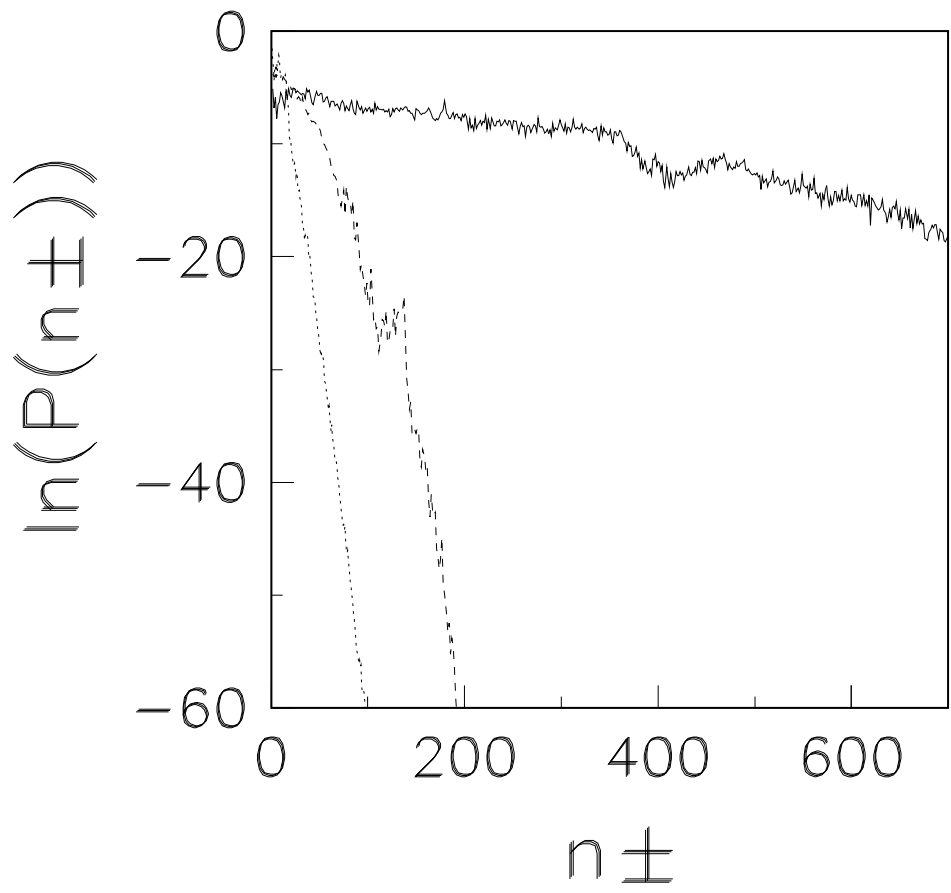
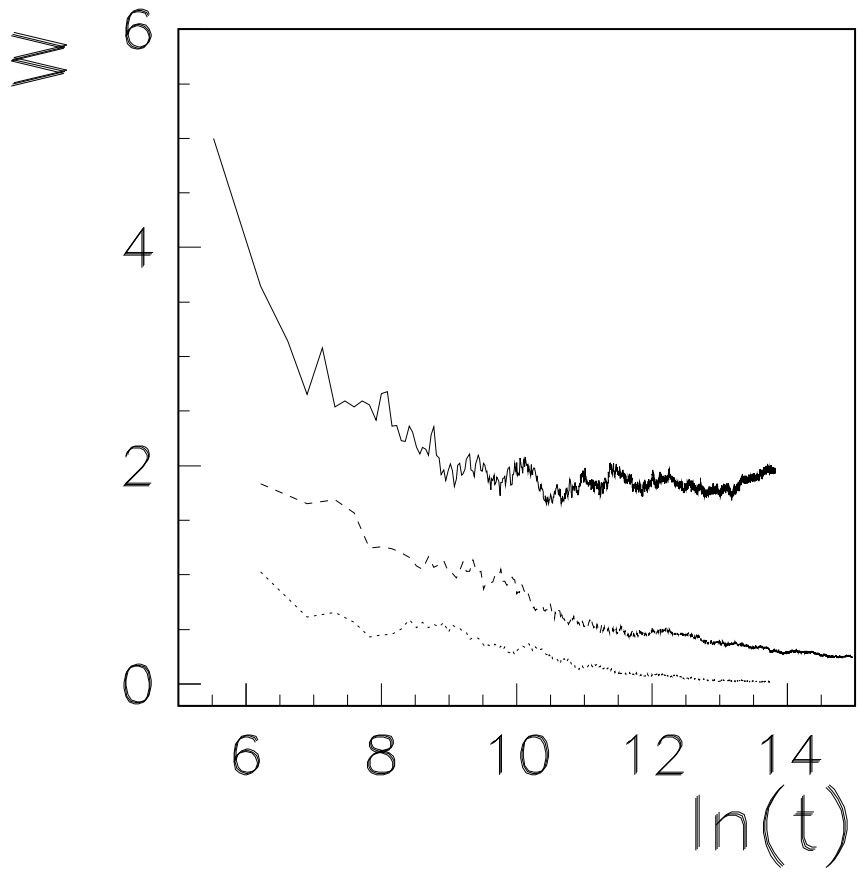
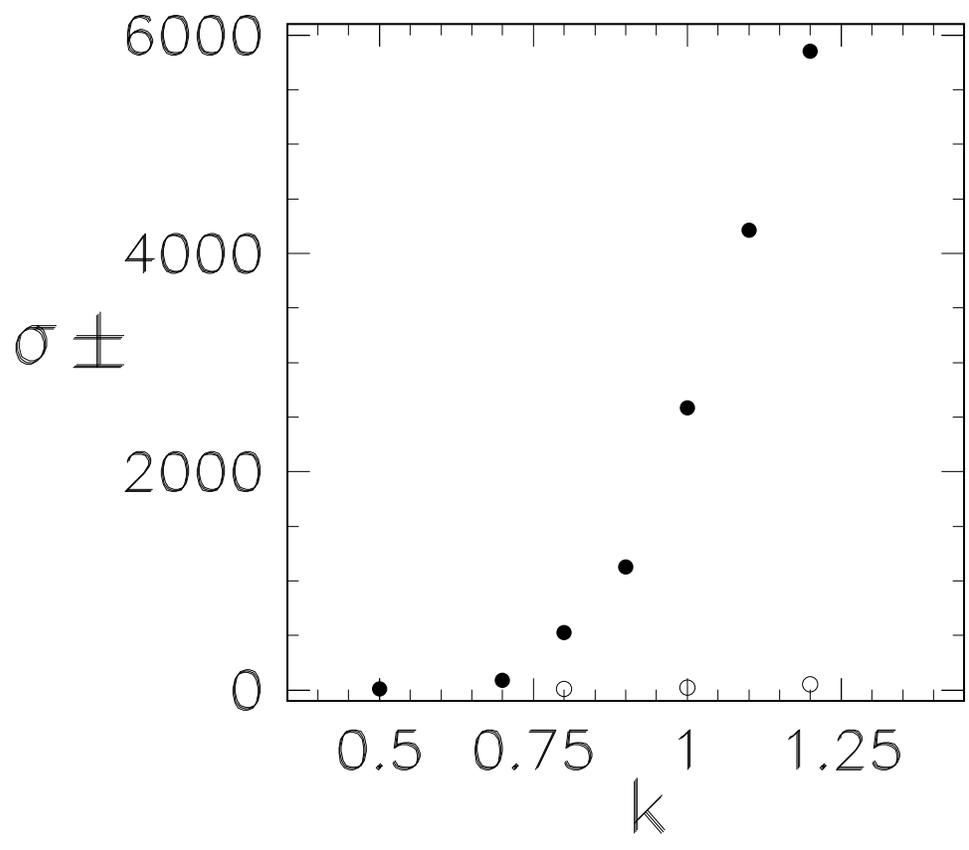
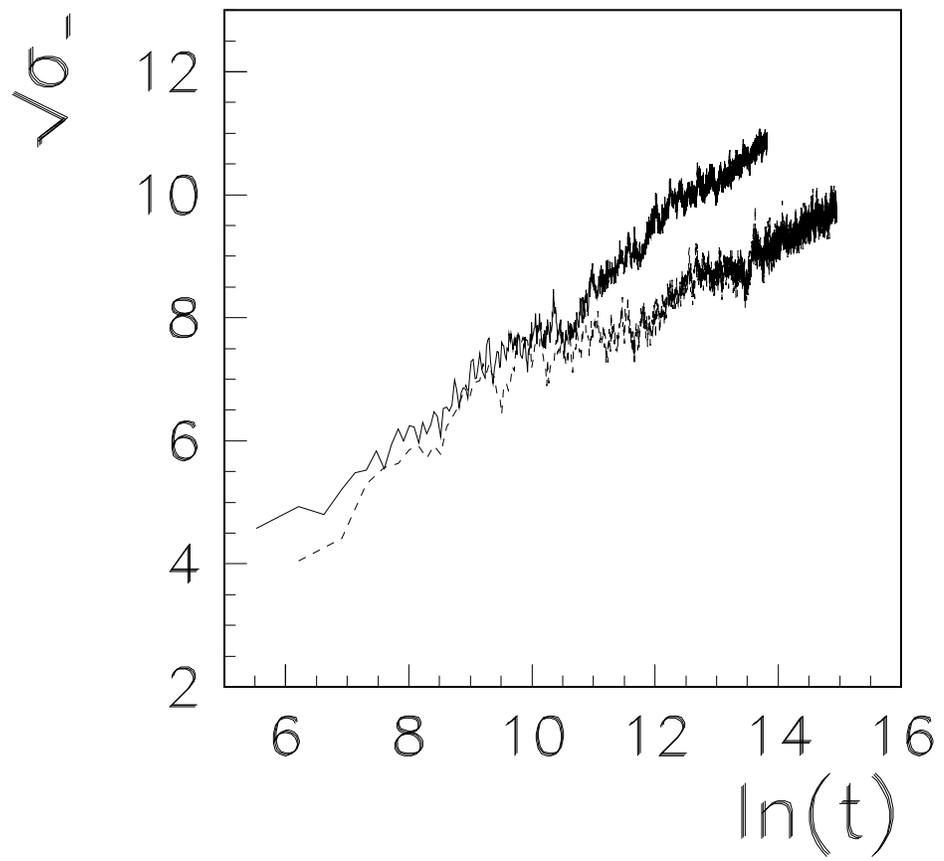
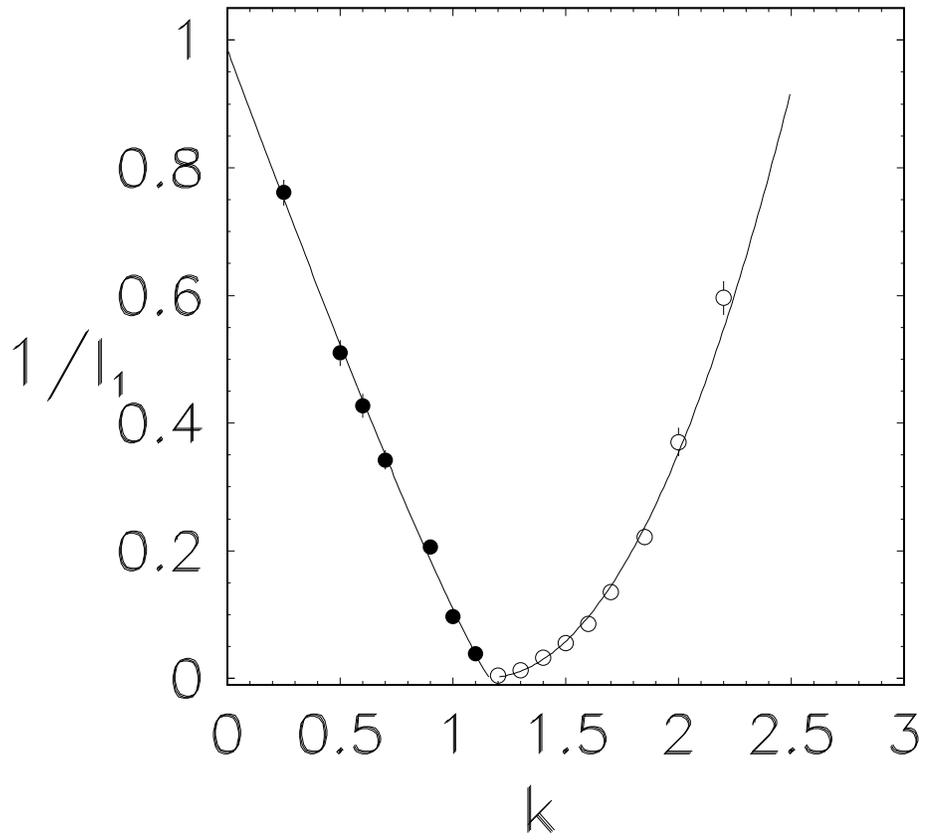
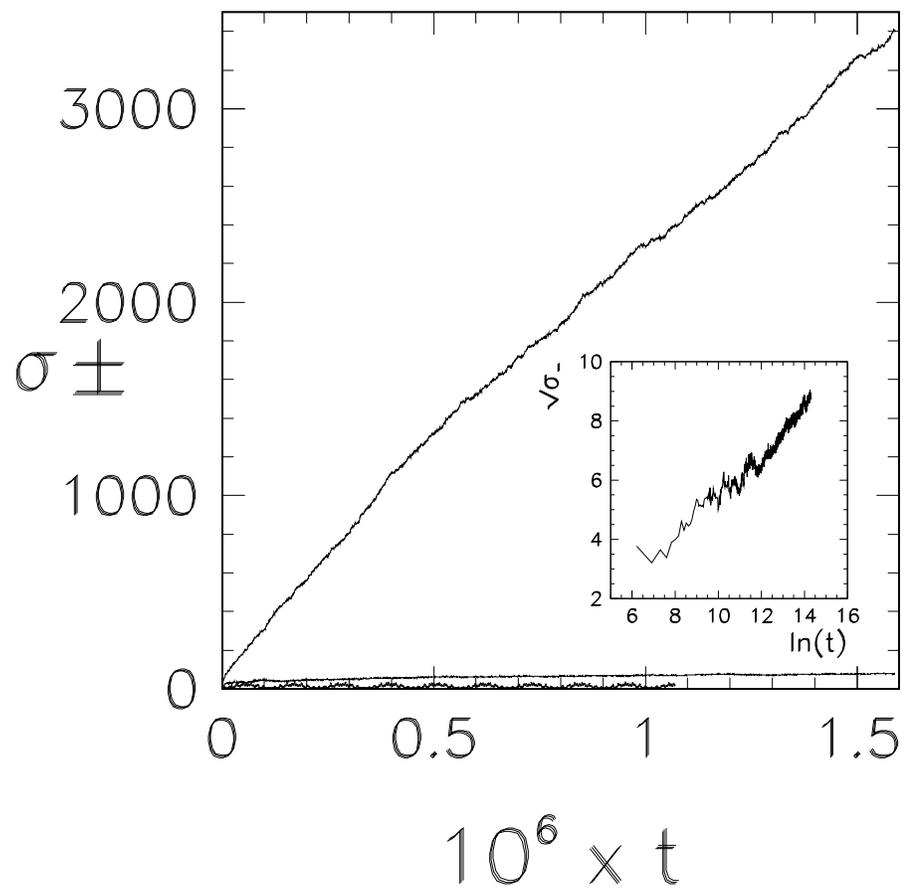
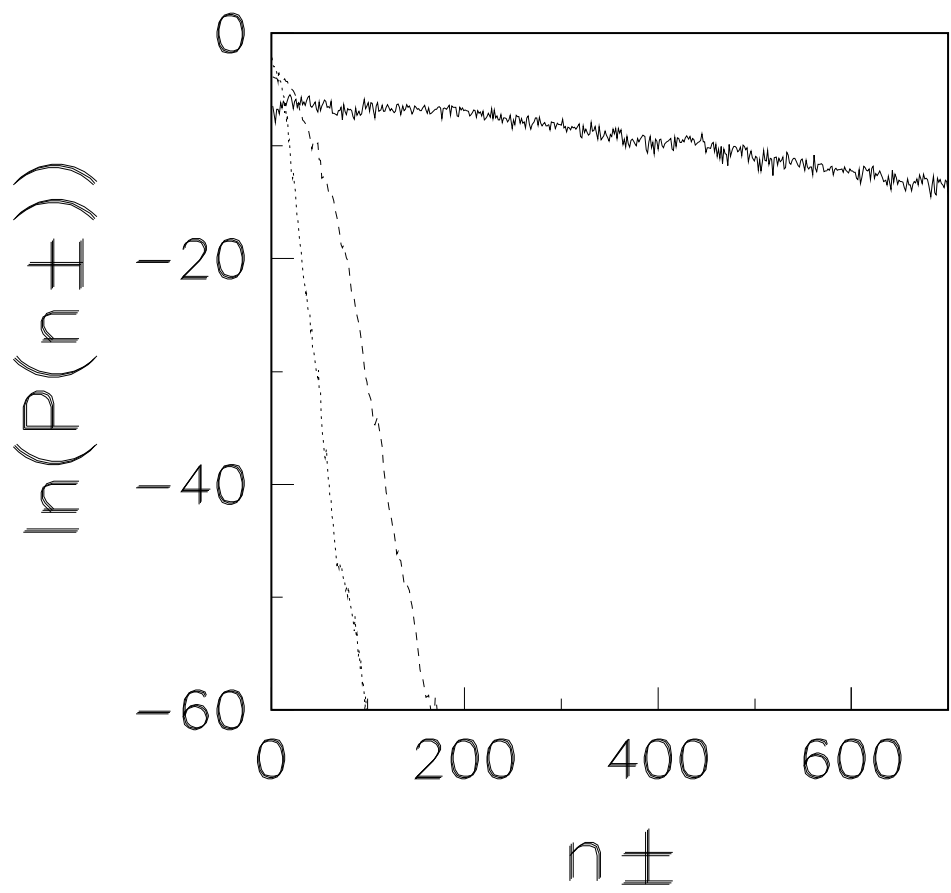
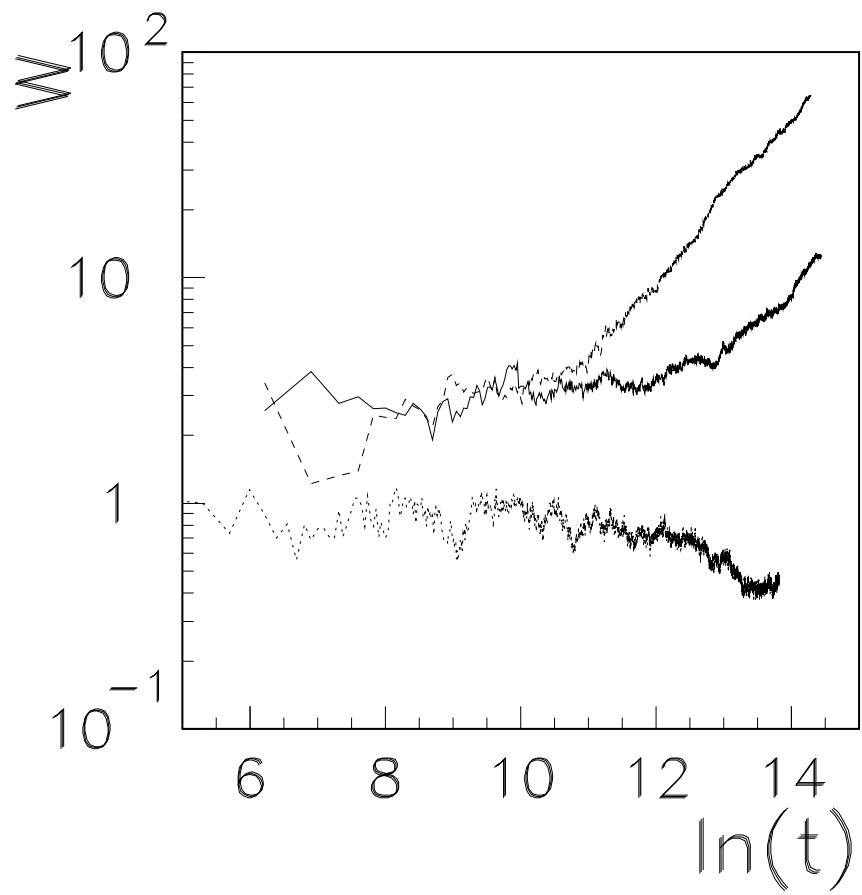
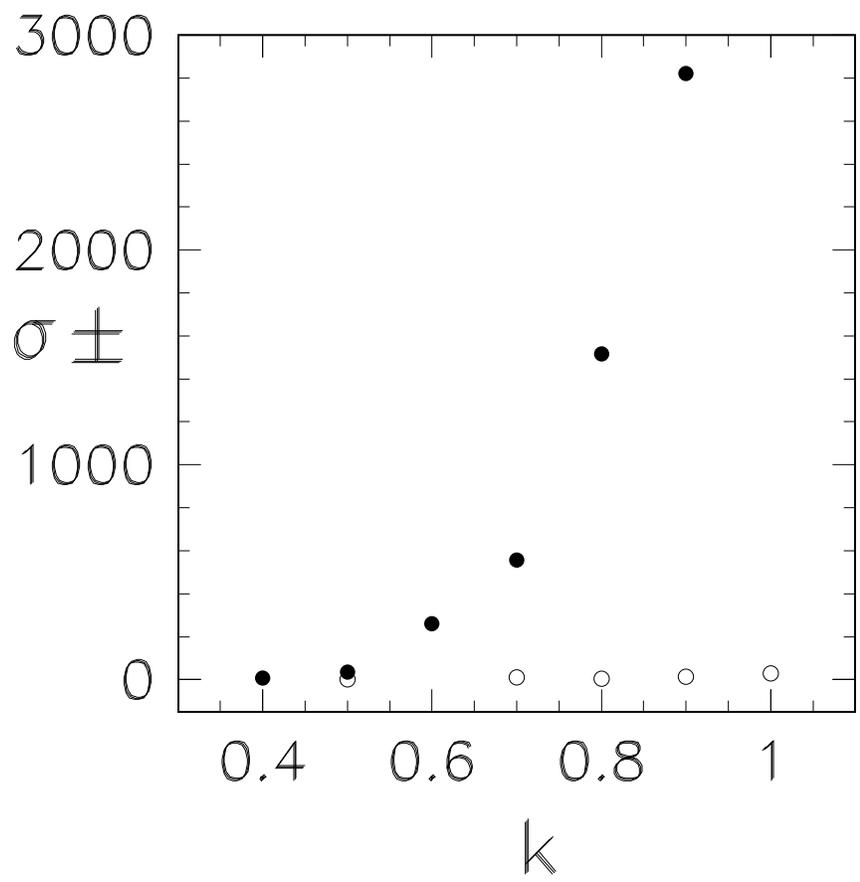
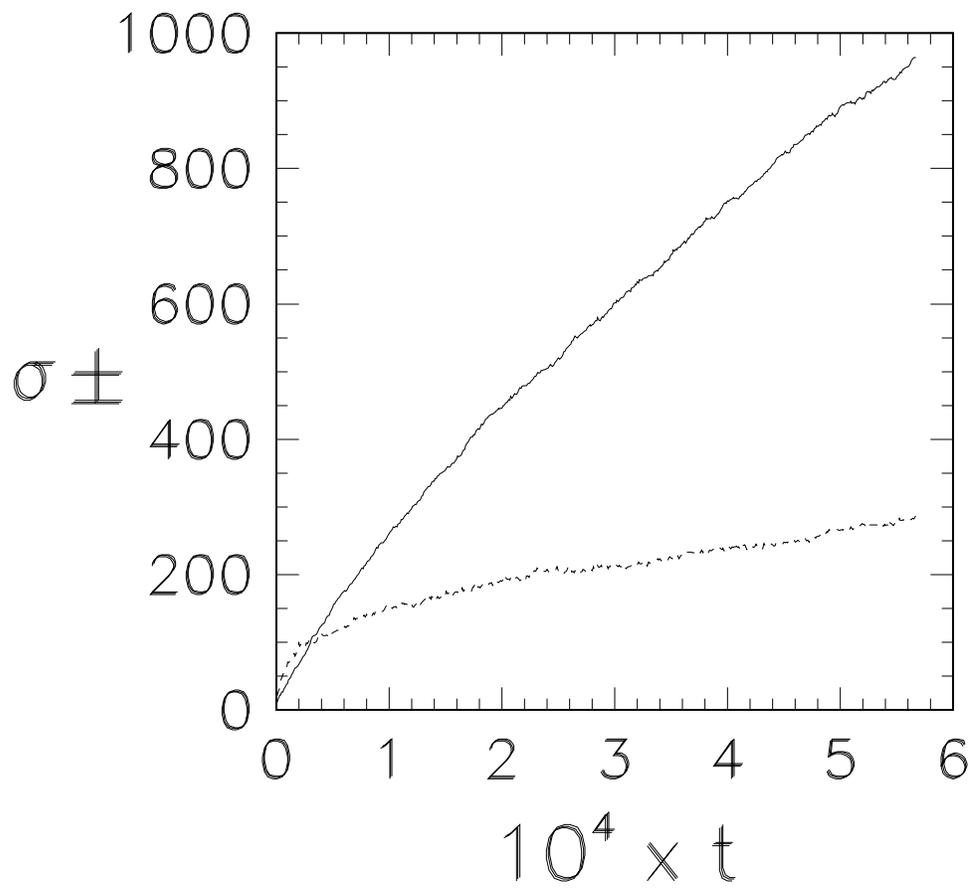
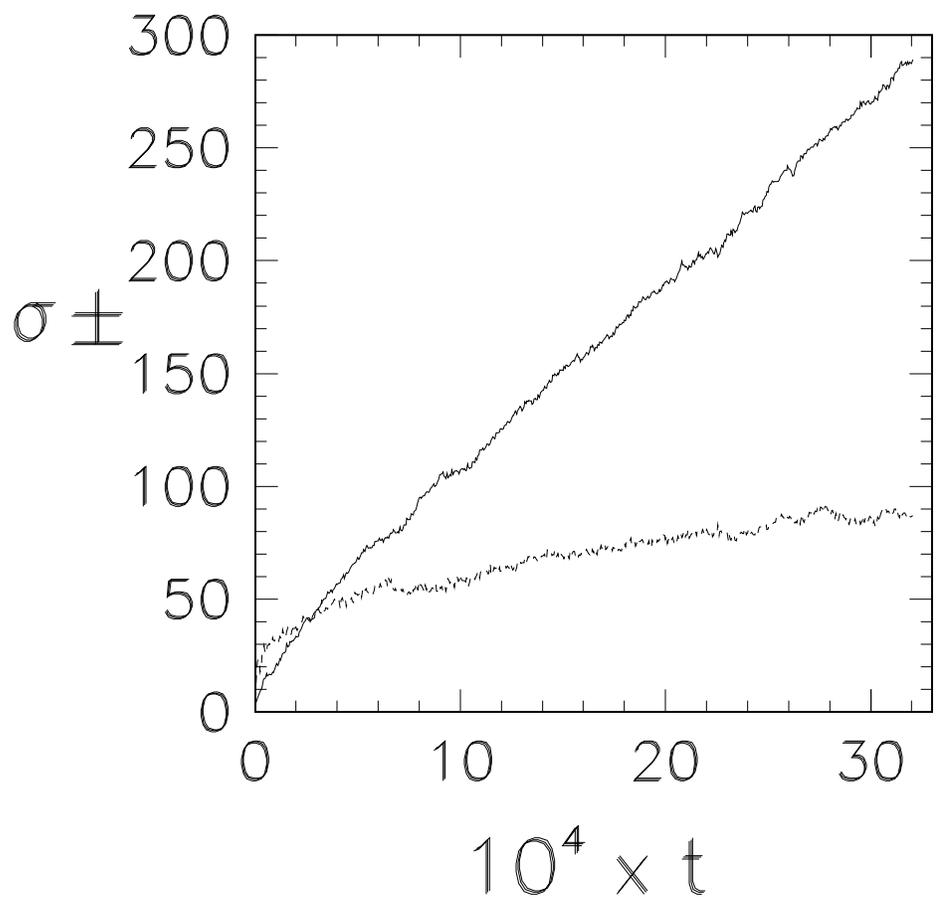
\end{document}